\documentstyle[psfig,sf99proc]{article}
\def\etal{\emph{et al.\ }}
\def\mj{$M_J$\ }

\title{Towards Understanding Jovian Planet Migration}

\authors{
Nelson, Andrew F. \affilmark{1}
and
Benz, Willy\affilmark{2}
}

\affiltexts{
\affiltext{1}{Max Planck Institute for Astronomy, Heidelberg, Germany},
\affiltext{2}{Universit\"at Bern, Bern, Switzerland}
}

\firstauthor
{AFN}
{andy@mpia-hd.mpg.de}

\authorsADS{%
Nelson, Andrew F.;
Benz, Willy;
}

\affiliationsADS{%
AA(Max Planck Institute for Astronomy )
AB(Universit\"at Bern )
}

\begin{document}

%
\begin{abstract}
We present 2D hydrodynamic simulations of circumstellar
disks around protostars using a `Piecewise Parabolic Method'
(PPM) code. We include a point mass embedded within the disk and
follow the migration of that point mass through the disk.
Companions with masses $M_c\ga 0.5M_J$ can open a gap in the 
disk sufficient to halt rapid migration through the disk. Lower
mass companions open gaps, but
migration continues because sufficient disk mass remains close
to the disk to exert large tidal torques. We find that the torques
which dominate the migration of low mass planets originate within a
radial region within 1--2 Hill radii of the planet's orbit radius,
a distance smaller than the thickness of the disk. We conclude
that a very high resolution 3D treatment will be required to 
adequately describe the planet's migration.

\end{abstract}

\section{Initial conditions}

The PPM code and initial conditions are very similar to those presented
in Nelson \etal 1998.  We begin with a one $M_\odot$ protostar fixed to
the origin of our coordinate system. We assume that a disk of mass
$M_D = 0.05 M_\odot$ is contained between the inner and outer grid
boundaries at 0.5~AU and 20~AU and that the disk is self gravitating.
A second point mass (the `planet') is set in a circular orbit
at a radius 5.2~AU away from the protostar and is free to migrate
through the disk in response to gravitational forces. No other
forces act on the planet and it does not accrete mass from the disk.
In different simulations, we investigate migration rates of different
planet masses.

The disk mass is distributed on a 128$\times$224 cylindrical ($r,\phi$)
grid with a surface density given by a power law,
$\Sigma(r) = \Sigma_1 \left({{1AU}/{r}}\right)^{p}$,
where $\Sigma_1$ is determined from the assumed disk mass and $p=3/2$.
We assume an initial temperature profile with a similar power law, 
$T(r) = T_1\left({{1AU}/{r}}\right)^{q}$,
where the temperature at 1~AU is $T_1=250$~K and $q=1/2$.  These
initial conditions produce a radial profile for which the minimum
Toomre $Q$ (of $\sim 5$) is found near the outer disk edge.
The profile exhibits a steep increase in the inner regions due to the
increased effects of pressure on the orbital characteristics there.
A single component isothermal gas equation of state is used to derive
pressure at each point in the disk.

Velocities are determined assuming initial rotational equilibrium. 
Radial velocities throughout are assumed equal to zero, while
angular velocities are determined by balancing the gravitational,
pressure and centrifugal forces in the disk.

\section{Migration rates with varying planet mass}

\begin{figure}[ht] 
\begin{center}
\leavevmode\psfig{file=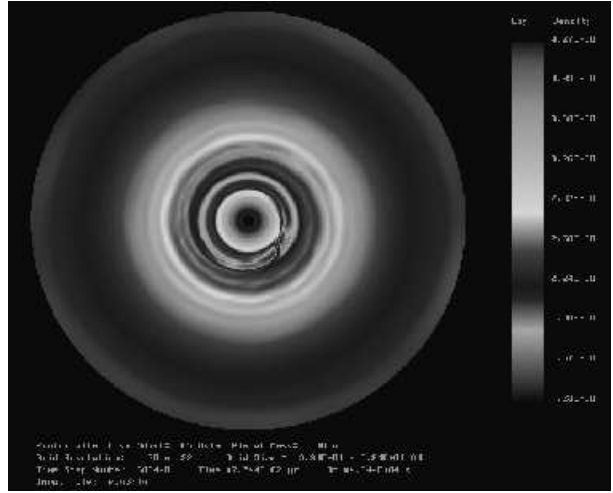,height=65mm,width=80mm,rheight=60mm,angle=-90}
\end{center}
\caption{
The surface density of the disk after 775 yr of evolution.
As shown by the deep blue region in the plot, the planet has built
a deep gap in the disk, but some spiral structure remains present in 
the gap. The planet has only just stopped its rapid migration through 
the disk at this time. }
\label{dens}
\end{figure}

Using the initial conditions outlined above we have completed a series
of simulations, varying the planet mass assumed for each simulation.
We show the effect of a 1\mj mass planet on the disk in figure \ref{dens}.
Within a few hundred years, the planet raises very large amplitude spiral
structures which lead (trail) the planet radially inside (outside) its
orbit radius and cause a gap to form around the planet. These structures are
similar to those in previous work (Bryden \etal 1999, Kley 1999) where
the planet's trajectory was fixed to a single orbit radius.

\begin{figure}[ht] 
\begin{center}
\leavevmode\psfig{file=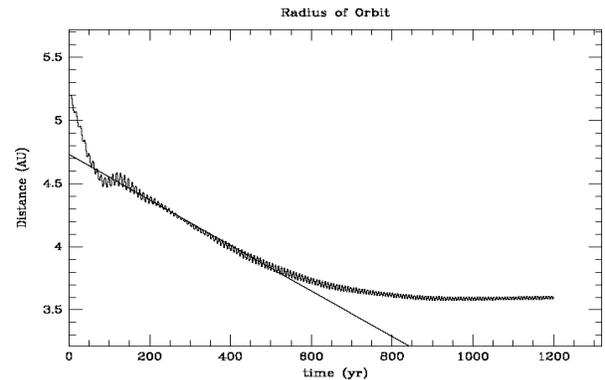,height=50mm,width=80mm,rheight=45mm,angle=-90}
\end{center}
\caption{
The orbital radius of the 1\mj planet shown in figure \ref{dens} as a
function of time. The planet begins at 5.2~AU and migrates inward to
$\sim3.5$~AU, before its inward trajectory is affected by the formation of
a sufficiently deep and wide gap. }
\label{traj}
\end{figure}

\begin{figure}[ht] 
\begin{center}
\leavevmode\psfig{file=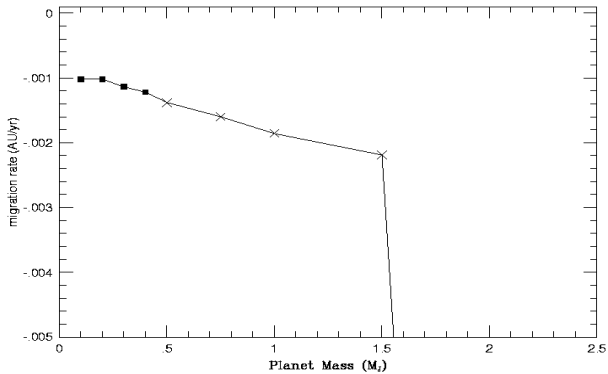,height=50mm,width=80mm,rheight=45mm,angle=-90}
\end{center}
\caption{
Migration rates of planets of varying mass.  Points marked `$\times$' are
able to evacuate a gap large enough and deep enough to halt their migration.
Planets greater than 1.5\mj `outrun' their own gap formation efforts and 
rapidly migrate inward to the inner grid boundary. }
\label{migrates}
\end{figure}

The orbital trajectory of the planet (figure \ref{traj}) is strongly affected
by the gravitational torques from the spiral structures. We show the
After a $\sim 50-100$~yr period in which the planet first builds spiral
structures, the migration rate is constant for the next 500~yr, then drops to 
zero and the planet reaches a `final' orbit radius of about 3.5 AU, after 
800~yr. The migration rates fit for simulations with different disk masses are
shown in figure \ref{migrates}, and are valid for the period for which
the migration rate is constant (see figure \ref{traj}), i.e. the period
for which the migration can be considered `Type~I'. The fitted migration
rates are very rapid: about 1~AU per thousand years, so that migration in to
the stellar surface would occur before the end of the disk lifetime of
10$^6$ yr. Still higher mass planets move so quickly through the disk that 
they `outrun' their own gap formation efforts and fall into the star.  With
more realistic initial conditions (a 2\mj planet should already have a gap), 
we expect this phenomenon to go away.

\begin{figure}[ht] 
\begin{center}
\leavevmode\psfig{file=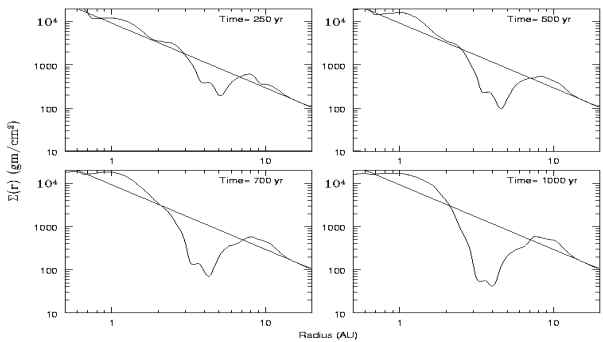,height=45mm,width=80mm,rheight=40mm,angle=-90}
\end{center}
\caption{Azimuth averaged density structure at several times during the
evolution. }
\label{azave}
\end{figure}

Over the course of the first several hundred years of evolution,
planets with mass less than 2\mj hollow out a deep gap in the disk,
which extends all the way around the star. In figure \ref{azave} we show 
the azimuth averaged surface density structure at several points during
the evolution of the system shown above. The gap forms quickly after
the beginning of the simulation and within 500 years has hollowed out
a region about 3~AU wide. The surface density near the planet (200--300~gm/cm$^2$) 
is a factor ten below the initial profile and less (100 gm/cm$^2$) at
its deepest, just outside the planet's orbit. By the end of the simulation,
the gap has deepened to a factor of 100 less than its unperturbed profile
and continued to get deeper even at the end of our simulations. It 
does not substantially increase its width after initial formation however.

The existence of a gap eventually causes the migration to slow and,
if it becomes deep and wide enough, decrease to a rate defined by the viscosity
of the disk (`Type II' migration) rather than to dynamical processes like
gravitational torques (`Type I' migration). From figure \ref{azave} we can 
determine the approximate disk conditions which define the transition between
Type~I and Type~II migration. From figure \ref{traj} we see that the
migration rate decreases starting after about 5-600 yr of evolution. Comparison
with figure \ref{azave} shows that the onset of the transition occurs when the 
gap is 3~AU wide and has surface density $\Sigma_{gap}\sim200-300$gm/cm$^2$.
The transition is concluded and further rapid orbital decay is suppressed
by the time the system has evolved for 700~yr, when the surface density
is $\Sigma_{gap}\sim 100$~gm/cm$^2$. 

\vspace{-2mm}

The values of surface density and gap width required for the onset
and completion of the transition to Type~II migration are typical of
of each of the simulations we have performed, withouth regard to planet mass.
Simulations with 0.3--0.5\mj planets are able to enter the transition
to Type~II migration but over the duration of our simulations ($1800$ yr) do
not complete the transition. We continue to evolve these simulations further
in time to determine their ultimate fate.

\section{The relative importance of the disk close to the planet}

Linear theory (e.g. Takeuchi \etal 1996) predicts that the most important
Fourier components of the spiral patterns raised in the disk will
be those with azimuthal wavenumber m=20--40, corresponding to an azimuthal
wavelength near the planet of about 1~AU. The Lindblad resonances of these
patterns will be at a distance radially inward and outward from the planet of 
$R_{LR} = a (1 \pm 1/m)^{2/3}\approx0.3$~AU, where $a$ is the 
semi-major axis of the planet. Another relevant parameter 
is the Hill radius, $R_H= a( M_J/{3M_*})^{1/3}\approx0.3$~AU, defining 
the sphere of influence of the planet. Further, the $z$ structure
of the disk becomes important on the same spatial scales because in 2D 
the gravity will be effectively `amplified' by the assumption that
all the disk matter is in the $z=0$ plane rather than some at high altitudes 
more distant from the planet. Unfortunately each of these values are
similar to the grid resolution size scale of computationally affordable
simulations.  

\begin{figure}[ht] 
\begin{center}
\leavevmode\psfig{file=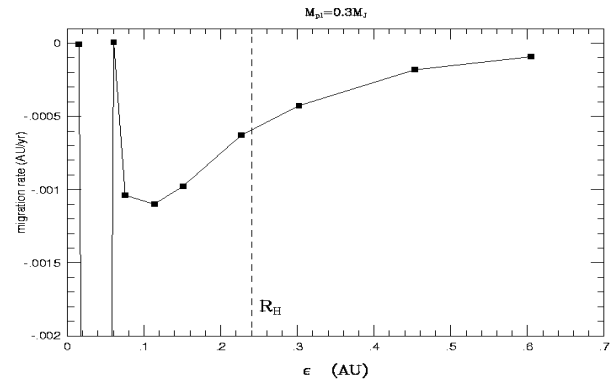,height=50mm,width=80mm,rheight=45mm,angle=-90}
\end{center}
\caption{The migration rate of a 0.3\mj planet with varying assumed 
gravitational softening length. The planet's Hill radius is shown with
a vertical dashed line. From the plot it is clear that the most influenetial
portion of the disk for migration is within 1--2 Hill radii of the planet.}
\label{sofrates}
\end{figure}

In order to characterize both physical and numerical effects we have performed
a series of simulations in which we vary the effective gravitational softening
length of the planet, effectively `turning on' or off the effect of matter
very close to the planet. The migration rates obtained from such a series of 
simulations are shown in figure \ref{sofrates}. We again assume the initial 
conditions as above, with a planet of mass 0.3\mj. 

\vspace{-3mm}

The migration rate increases by a factor of about five as the softening
decreases from 0.5~AU to 0.1~AU, clearly showing the importance of the
distribution of disk matter close to the planet.  The largest increases
occur as $\epsilon$ decreases to the size of the Hill radius or smaller.
Below $\epsilon=0.08$~AU, or half the size of one grid cell, the migration
rate is numerically unstable and slight changes in the softening change the
migration rate from zero to 1 AU/100~yr (below the bottom of the plot).
The dependence of the migration on the disk matter very near the planet
is in qualitative agreement with with the conclusion of Ward (1997), who
showed that the most important contribution to the migration will come from
the disk matter within about one disk scale height (in our case $\sim 0.3$~AU
near the planet) of the planet's radial position. These results show that very
high resolution three dimensional simulations of the region around the planet
will be required in order to understand the migration rates of a planet
through the disk.

\section{Summary}

In the course of this study we have

$\bullet$ Shown that planets evolved in a 2D disk without a gap
migrate inward through a substantial fraction of their initial
semi-major axis radius on timescales $\sim 1000$ yr.

$\bullet$ Shown that planets with masses higher than $\sim 0.5$\mj can
open a gap sufficiently wide and deep to drastically slow their migration
through the disk (i.e. transition to Type~II migration). Conditions for
beginning the transition are that surface density of the disk near the planet
be about 200-200 gm/cm$^2$ and that the gap be $\sim 3$~AU wide. Conditions
for completing the transition to Type~II migration are surface densities 
near the disk of $\sim 100$ gm/cm$^2$.

$\bullet$  Demostrated the critical importance of very high spatial
resolution of the disk near the planet required for correct evolution
of the planet's migration.

\section*{References}

\reference
Bryden, G., Chen, X., Lin, D. N. C., Nelson, R. P., Papaloizou, J. C. B.
1999, ApJ, 514, 244

\reference Kley, W., 1999, MNRAS, 303, 696

\reference Nelson, A. F., Benz, W., Adams, F. C., Arnett. W. D., 1999, ApJ,
502, 342

\reference Takeuchi, T., Miyama, S. M., Lin, D. N. C., 1996, ApJ, 460, 832

\reference Ward, W., 1997, Icarus, 126, 261

\vspace{-1mm} 

\end{document}